\documentclass[twocolumn,showpacs,floatfix,10pt,nofootinbib]{revtex4}
\pdfoutput=1
\usepackage{epsfig}\usepackage{float}
\RequirePackage{amssymb}
\usepackage{makeidx}
\usepackage[latin1]{inputenc}
\usepackage{graphicx}
\usepackage{indentfirst}
\usepackage{color}
\newcommand{\be}{\begin{equation}}
\newcommand{\ee}{\end{equation}}
\newcommand{\ud}{\mathrm{d}}

\newcommand{\ba}{\begin{eqnarray}}
\newcommand{\ea}{\end{eqnarray}}

\newcommand{\el}{^}

\newcommand{\lp}{\ell_{\rm p}}
\newcommand{\mpl}{M_{\rm p}}

\newcommand{\md}{M_{(5)}}

\newcommand{\ld}{\ell_{(5)}}
\newcommand{\mew}{M_{\rm ew}}

\begin{document}

\title{Effective Monopoles within Thick Branes}

\author{J. M. Hoff da Silva}
\email{hoff@feg.unesp.br;hoff@ift.unesp.br}
\affiliation{Departamento de F\1sica e Qu\1mica, Universidade
Estadual Paulista, Av. Dr. Ariberto Pereira da Cunha, 333, Guaratinguet\'a, SP,
Brazil.}

\author{Rold\~ao da Rocha}
\email{roldao.rocha@ufabc.edu.br}
\affiliation{Centro de Matem\'atica, Computa\c c\~ao e Cogni\c c\~ao, Universidade Federal do ABC 09210-170, Santo Andr\'e, SP, Brazil.}

\pacs{11.25.-w, 03.50.-z, 14.80.Hv}

\begin{abstract}
The monopole mass is revealed to be considerably modified in the thick braneworld paradigm, and depends on the position of the monopole in the brane as well. Accordingly, the monopole radius continuously increases, leading to an unacceptable setting that can be circumvented when  the brane thickness has an upper limit.
Despite such peculiar behavior, the  quantum corrections accrued --- involving the classical monopole solution --- are shown to be still under control. {We analyze the monopole's peculiarities also taking into account the localization of the gauge fields.} Furthermore, some additional analysis in the thick braneworld context and the similar behavior evinced by the topological string are investigated.\end{abstract}

\maketitle

\section{Introduction}

The solution of the mass hierarchy problem has acquired  a new perspective since 1999, from the braneworld viewpoint. Indeed, it is straightforward to realize in the Randall-Sundrum  model (RSI) that the warped non-factorizable geometry is responsible to dress the naked mass parameters via the Higgs mechanism \cite{RSI}. Such models are devoid of the necessity of any hierarchy to reproduce TeV mass scales from the fundamental ($\sim$ $10\el{19}$ GeV) scale. Notwithstanding, in the RSI model our world is described by an infinitely thin  three-brane
endowed with a local coordinate chart $x^\mu$, {\footnotesize{$\mu = 0,1,2,3$}},
embedded in a five-dimensional bulk with the metric
\be
\ud s^2
=
e^{-\sigma\,|r|}\,g_{\mu\nu}\,\ud x^\mu\,\ud x^\nu
+ \ud r^2.
\label{g5}
\ee
Here $r$ denotes the fifth dimension and
$\sigma^{-1}$ is a length determined by the brane tension.
This parameter relates the four-dimensional Planck mass
$\mpl$ to the five-dimensional gravitational mass $\md \simeq 1\,$TeV$/c^2$.
Perceive that experimental limits require $\md\gtrsim 1\,$TeV,
but there is no strong theoretical evidence that places
$\md$ at any specific value below $\mpl$.

The RSI model may be realized as an approximate compactification scheme,  taking into account the existence of a fundamental length below which classical physics does not hold. The brane must, therewith, present a thickness $\Delta$, where deviations from the four-dimensional Newton's law
occur in such scales. Besides, it is well known that thick --- instead of thin --- branes may circumvent some problems as, e. g.,  the proton decay (see the Appendix for a discussion). Current precision experiments require that
$\Delta \lesssim 44\,\mu$m~\cite{kapp},
whereas theoretical reasons imply that
$\Delta\gtrsim\ld\approxeq \lp\,\mpl/\md \approx 2.0 \times 10^{-19}\,$m.
Hereon  $1=c=\hbar=\mpl\,\lp=\ld\,\md$,
where $\mpl\simeq 2.2\times 10^{-8}\,$kg and
$\lp\simeq 1.6\times 10^{-35}\,$m
are respectively the Planck mass and the length related to the
four-dimensional Newton constant $G_{\rm N}=\lp/\mpl$.
In our analysis the five-dimensional scenario
with $\md\simeq \mew\simeq 1\,$TeV $\approx 1.8\times 10^{-24}\,$kg is considered,
the electroweak scale corresponding to the length
$\ld\simeq 2.0\times 10^{-19}$m. The first step towards a more realistic scenario was accomplished quite soon after the RSI model \cite{MG, OT}, in the consideration of a scalar field coupled to gravity. The solution propounded in \cite{MG} presents full compliance with a (thick) domain wall, interpolating between two asymptotic five-dimensional anti-de Sitter  spaces. In this context, the brane --- performed by the domain wall --- is not a delta like object, and the brane thickness is encrypted in two parameters,  which composes the superpotential. (For an up to date review on thick brane solutions see, e.g., \cite{VVM}). It is common in the literature (see e.  g. \cite{MG}) to fix one of such parameters to be constant, and to analyze the brane thickness with respect to the another floating parameter on the selected superpotential. In our approach here, based on the experimental and theoretical reasons indicated in the previous paragraph, one parameter is not fixed \emph{a priori}, but the parameter space of such constrained parameters is depicted and analyzed. Since the brane performs our observed universe, it is mandatory and quite conceivable to further explore what are the physical implications of considering topological defects in the warped domain wall. We are going to delve into a more detailed and deep analysis hereon.

A topological defect is a classical and stable solution appearing in abelian as well as non abelian systems. Furthermore, it presents internal structure, which can be affected by the warped geometry. In this paper we shall study the typical toy model for monopoles --- namely, the one arising from a $SO(3)$ symmetry breaking --- living in a warped thick braneworld. The typical monopole radius is shown to be affected by the warped geometry, in an analysis regarding the effective monopole action on the brane. Obviously it is not a consequence of any relativistic dynamics. Indeed, as further investigated in more details, this effect may be outlined  in what follows. Consider the radii of two monopoles at rest, in the thick brane scenario. One of them (assume the monopole $A$) is located at the brane core, and another monopole $B$ is placed at some proper distance, out the brane core. The monopole $B$ radius is shown to be bigger than the monopole $A$ radius. The reason for such a behavior is an aftermath of the fact that the (naked) monopole radius is a dressed parameter likewise, when considered within a warped brane. It is forthwith realized, since the monopole radius scales as the inverse of the monopole mass, and the mass parameter is dressed by the warp factor, as the standard analysis of \cite{RSI}. More specifically, the mass parameter is dressed via the spontaneous symmetry breaking parameter in a Higgs-like potential. As the dressed (`visible') radius is dependent on the monopole position within the brane, one may straightforwardly impose an upper limit on the brane thickness. Hence, the study of such classical solution may impose some constraints improving the braneworld model concerned. The upper limit constrains, under experiment-based conditions on the brane thickness, two parameters in the proposed model, contrary to approaches in the literature hitherto, that fix one of the parameters instead \cite{MG}.

Let us say a few words about our procedure. In the following analysis we consider, as mentioned, effective monopoles. By `effective' we mean monopoles whose fields do not retroact with the gravitational background and survive to the damping mechanism in the monopole creation caused by, for instance, the inflationary scenario. Of course, a comprehensive analysis involving the monopole rate production in the braneworld cosmology context would be interesting, but it is obviously a hard task. The approach we shall consider, instead, the (survivor) monopoles already trapped on the thick brane. In other words, we are assuming that the braneworld cosmology is not widely different, in which concerns the existence of the monopoles, from the usual cosmological scenario. Certainly it is incomplete, but at the same time a rather conservative approach, since we do not want to depart from the standard cosmological model results. {In addition, we study some aspects of the monopole, taking into account the gauge field localization on the brane.}

This paper is structured as follows: in the next Section the 't Hooft-Polyakov magnetic monopole within a thick and warped (unspecified) domain wall is investigated, and the general physical implications are analyzed. The technical details are straightforward. In Section III,  the monopoles are regarded and studied in a specific background.  The final Section is devoted to conclude and provide prominent outlooks, pointing out some essential remarks. We leave for the Appendix a heuristic argumentation in favor of the existence of a brane thickness, as well as the study of how the gauge field localization issue affects some of the previous results.

\section{$SO(3)$ Monopoles in warped thick Braneworlds}

Consider the 't Hooft-Polyakov monopole configuration (of the scalar field), based upon an initially $SO(3)$ invariant Lagrangian \cite{TP}. Within a brane embedded in a warped spacetime, the calculations are indeed straightforward. The unique modification is elicited from the gravitational weight present in the invariant volume, necessary to encompass the effective Lagrangian in the warped space.

The background, which in this context is unaffected by the monopoles, is assumed to be endowed with the following warped line element:
\be ds\el{2}=e\el{2A(r)}\Bigg(dt\el{2}-\sum_{i=1}\el{3}dx_{i}\el{2}\Bigg)-dr\el{2},\label{1} \ee where $r$ denotes the extra dimension. The functional form of $A(r)$ depends on the model used to accomplish the braneworld scenario. In the next Section a specific example is considered. The effective Lagrangian reads
\ba S_{eff}&=&\int d\el{4}x \;\sqrt{\bar{g}}\;\Bigg\{-\frac{1}{4}\bar{g}\el{\mu\alpha}\bar{g}\el{\nu\beta}F_{\mu\nu}\el{a}F_{\alpha\beta}\el{a}\nonumber\\&&+\frac{1}{2}\bar{g}\el{\mu\nu}D_{\mu}\phi\el{a}D_{\nu}\phi\el{a}-\frac{\lambda}{8}\Big(\phi\el{a}\phi\el{a}-v\el{2}\Big)\el{2}\Bigg\},\label{2}\ea  where
\ba D_{\mu}\phi\el{a}&=&\partial_{\mu}\phi\el{a}-e\varepsilon\el{a}_{bc}W_{\mu}\el{b}\phi\el{c},\label{3} \\
F_{\mu\nu}\el{a}&=&\partial_{[\mu}W_{\nu]}\el{a}-e\varepsilon\el{a}_{bc}W_{\mu}\el{b}W_{\nu}\el{c} \label{4} \ea and $\bar{g}_{\mu\nu}=g_{\mu\nu}(x\el{\mu},r=\bar{r})$, where $\bar{r}$ denotes the brane core localization. The spacetime indices run from $0$ to $3$, and the internal indices are denoted by $a,b,c$. Taking into account Eq. (\ref{1}), it follows that
 \ba S_{eff}&=&\int d\el{4}x\; \Bigg\{-\frac{1}{4}F_{\mu\nu}\el{a}F\el{a\mu\nu}+\frac{1}{2}e\el{2A}D_{\mu}\phi\el{a}D\el{\mu}\phi\el{a}\nonumber\\&&-\frac{\lambda}{8} e\el{4A}\Big(\phi\el{a}\phi\el{a}-v\el{2}\Big)\el{2}\Bigg\}.\label{5}\ea After a wave function renormalization  $\phi\el{a}\mapsto e\el{-A}\phi\el{a}$,  just a $SO(3)$ version of the analysis performed in the RSI model, the effective action $S_{eff}$ reads \ba \hspace*{-0.05cm}\int d\el{4}x \Bigg\{-\frac{1}{4}F_{\mu\nu}\el{a}F\el{a\mu\nu}+\frac{1}{2}D_{\mu}\phi\el{a}D\el{\mu}\phi\el{a}-\frac{\lambda}{8}\Big(\phi\el{a}\phi\el{a}-\tilde{v}\el{2}\Big)\el{2}\Bigg\},\label{6}\nonumber\ea where $\tilde{v}$ denotes the dressed spontaneous symmetry breaking parameter, related to the associated naked parameter by $\tilde{v}=ve\el{A}$. The effective Lagrangian defined by Eq. (\ref{5}) provides a perturbative spectrum quite usual for the $SO(3)$ models via Higgs mechanism, but with the dressed parameter entering in the masses. Herewith, the particle content includes a massless photon ($W_{\mu}\el{3}$) coupling to the unbroken $U(1)$ current and to the vector bosons $W_{\mu}\el{\pm}$, and a neutral scalar as well. The masses are given by \be M_{W}=e\tilde{v} \label{7}\ee  and \be M_{H}=\sqrt{\lambda}\tilde{v},\label{8} \ee respectively. It is well known that, beyond the perturbation theory, there is a stable and time-independent solution that behaves as a particle in the classical theory \cite{TP}. Furthermore, the long range gauge field of this solution matches the field of a Dirac magnetic monopole \cite{DIRAC}. This solution is nothing but the magnetic monopole. The explicit construction of the monopole solution  is already part of several textbooks in QFT, and it is not aimed to be repeated here. It is interesting, notwithstanding, to address to some peculiar points of prior importance throughout the paper, regarding the monopole configuration.

The scalar and vector field configurations for the monopole solution read, respectively \ba \phi\el{a} ({\bf \rho})&=&\tilde{v}H(M_{W}\rho)\;\hat{r}^{a}\label{9}\\
 W_{b}^{a} ({\bf \rho})&=&\frac{\varepsilon_{bac}}{e\rho}[1-G(M_{W}\rho)]\hat{r}^{c},\label{10}\ea where $H$ and $G$ are radial functions which minimize the energy associated to the monopole\footnote{Numerical solutions for $H$ and $G$ may be found in Refs. \cite{NS,NSN}.}. The appropriate boundary conditions imposed by the finite mass requirement are $H=0$ and $G=1$ as $\rho\rightarrow 0$; and $H=1$ and $G=0$ as $\rho\rightarrow \infty$. The monopole mass, given by
\be M_{m}=\frac{4\pi}{e\el{2}}M_{W}C(\lambda/e\el{2}),\label{11}\ee  is strongly dependent on the vector bosons mass. The term $C(\lambda/e\el{2})$ is a numerical factor nearly independent of $\lambda/e\el{2}$ \cite{NSN}. Eq. (\ref{7}) implies that the warp factor dressing the parameter $v$ is felt likewise by the monopole mass living in the thick brane.

In the classical theory it is possible to determine a typical radius $R_{m}$ for the monopole \cite{JP}. The size $R_{m}$ is chosen to balance the energy stored in the magnetic field outside the monopole core ($\sim g\el{2}/R_{m}$, where $g=e\el{-1}$ is the monopole charge), and the energy due to the scalar field gradient in the monopole core ($\sim M_{W}\el{2}R_{m}$). Hence  $R_{m}\sim M_{W}\el{-1},$  in order of magnitude. Now, taking into account Eq. (\ref{7}) and the dressed parameter $\tilde{v}$, the monopole radius may be recast in the following straightforward form
\be  \tilde{R}_{m}\sim R_{m}e\el{-A}, \label{12} \ee where $R_{m}$ denotes the naked radius ($\sim 1/ev$).  Before studying the physical implications of dressing the monopoles mass and radius, let us stress a point of prominent  importance, concerning the behavior of these parameters along the brane thickness. First, note that the monopole mass may be written as well in terms of its naked counterparts by \be \tilde{M}_{m}=M_{m}e\el{A},\label{13} \ee where $M_{m}=4\pi v C/e$. Hence, although the mass and the radius are certainly affected by the monopole position in the thick brane, its product is invariant: $\tilde{R}_{m}\tilde{M}_{m}=R_{m}M_{m}$. This is a cogent argument to the fact that the relation $\lambda_{c}/R_{m}\ll 1$ is also preserved when dressed by the warp factor: $\tilde{\lambda}_{c}/\tilde{R}_{m}=\lambda_{c}/R_{m}$. Here $\lambda_{c}$  denotes the monopole Compton wavelength. Herewith, no matter what type of modification the monopole experiences in a thick brane, its quantum corrections are still under control.

\section{A specific background: The Gremm's Model}

In order to ascertain the physical implications regarding Eqs. (\ref{12}) and (\ref{13}),  the background is particularized henceforward. For the sake of completeness, some of the main steps necessary to the  implementation of the model presented in Ref. \cite{MG} are briefly introduced. As mentioned, this model is based on five-dimensional gravity coupled to a scalar field. The action reads
\be S=\int d\el{4}x\, dr\, \sqrt{g}\Bigg(\frac{1}{2}(\partial \Phi)\el{2}-V(\Phi) -\frac{R}{4}\Bigg),\label{14} \ee where $R$ denotes the scalar curvature, and the scalar field $\Phi$ depends only on the extra dimension. The equations of motion should supply the solution to the scalar field as well as the $A(r)$ function appearing in Eq. (\ref{1}). They are given by $\frac{d\Phi}{dr}=\frac{1}{2}\frac{\partial \mathbb{W}(\Phi)}{\partial \Phi}\equiv\frac{1}{2}\mathbb{W}_{\Phi}\label{15}$ {and}
$\frac{dA}{dr}=-\frac{1}{3}\mathbb{W}(\Phi),$ whenever the potential $V(\Phi)$ is written in terms of the superpotential $\mathbb{W}(\Phi)$ as \cite{SUP}
\be V(\Phi)=\frac{1}{8}\mathbb{W}_{\Phi}\el{2}-\frac{1}{3}\mathbb{W}(\Phi)\el{2}.\label{17}\ee The selected superpotential of Ref. \cite{MG} is chosen as \be \mathbb{W}(\Phi)=3bc \sin\left(\sqrt{{2}/{3b}}\;\Phi\right). \label{18}\ee Consequently, the solution reads \be \Phi(r)=\sqrt{6b} \arctan\Big(\tanh\Big({cr}/{2} \Big)\Big)\label{19} \ee and \be A(r)=-b \ln(2\cosh(cr)),\label{20} \ee where the parameter $c$ is related to the brane thickness and the composite parameter $bc$ provides the $AdS$ curvature in this model. In addition, the brane core is localized at $r=0$, corresponding to the point related to the maximum slope of the scalar field (performing the domain wall).

Returning to our problem,  although the 't Hooft-Polyakov monopole solution is well known  obtained with a $SO(3)$ initial gauge group (not present in the standard model), the existence of monopoles follows from the general scope of unification theories. In general grounds, a grand unified theory is endowed with a large group of exact gauge symmetries which are spontaneously broken at an extremely large mass scale. Such a mass scale depends on the grand unified model dealt with, and it is estimated to be of order $10\el{14}$ GeV \cite{WEI}. Hence the naked mass and radius expected for the monopole are given by $10\el{16}$ GeV and $10\el{-28}$ cm, respectively.

In view of Eqs. (\ref{20}) and (\ref{12}), it follows that \be \tilde{R}_{m}\sim R_{m}2\el{b}[\cosh(cr)]\el{b}.\label{21} \ee The point to be stressed here is that, on the brane core, the radius --- in this example given by $R_{m}2\el{b}$ --- has order $10\el{-28}$ cm. Nevertheless, the monopole radius depends on the extra dimension in the (thick) brane,  via the warp factor. Thereby, the monopole radius becomes bigger outside the brane core\footnote{One may consider the possibility of redefining this criterium by taking the value $10\el{-28}$ cm at the brane surface. However, this situation is complicated, since at the brane core the radius shall be much less than $10\el{-28}$cm, i. e., one basically faces the problem of a point monopole which is essentially a strong-coupling regime; we just cannot calculate anything.}. Therefore,  to avoid (unobserved) monopoles with mass scale of order TeV, taking into account that $R_{m}\sim 1/M_{m}$,  the condition $R_{m}2\el{b}[\cosh(cr\el{*})]\el{b}\lesssim10\el{-15} {\rm cm}\label{22}$ must be imposed, where $r\el{*}$ denotes the brane `surface' along the extra dimension. Hence, it reads \be 2\el{b} \Bigg[\cosh\Big(\frac{c\Delta}{2}\Big)\Bigg]\el{b}\lesssim10\el{13},\label{23} \ee where $\Delta=2r\el{*}$ is defined to provide a naive account for the brane thickness. In the analysis of metric fluctuations, the particular case $b=1$ is studied \cite{MG}, leaving $c$ as the unique parameter setting the $AdS$ curvature as well as the brane thickness. In this case, the constraint encoded in Eq. (\ref{23}) implies approximately that $ \Delta\lesssim {60}/{c}$. Emphasizing that when the $AdS$ curvature is  infinity $c\rightarrow \infty$, and one recovers the thin wall limit, as expected in the case $b=1$ \cite{MG}. The peculiar behavior of magnetic monopoles in the thick wall provides, therefore, a boundary on the parameter controlling the brane thickness.

More generally, as the  brane thickness $\Delta$ must  satisfy $2.0 \times 10^{-19} $m $ \lesssim \Delta \lesssim 44\; \mu $m \cite{kapp}, based on such physical constraint it is not necessary to fix the parameter $b$. The most general relation between the parameters $b$ and $c$ is now analyzed. Eq. (\ref{23}) implies that \be \Delta \lesssim\frac{2}{c} \arccos{\hspace{-0.5mm}{\rm h}}\left(\frac{ 10^{13/b}}{2}\right).\label{b1} \ee\noindent As 44 $\mu$m is the physical supremum for the brane thickness above in (\ref{b1}) it follows that $\frac{2}{c} \arccos{\hspace{-0.5mm}{\rm h}}\left(\frac{ 10^{13/b}}{2}\right) \lesssim 44\,\mu{\rm m}$. Therefore, \be 1.0\times 10^{-19} c \lesssim \arccos{\hspace{-0.5mm}{\rm h}}\left(\frac{ 10^{13/b}}{2}\right) \lesssim 2.2 \times 10^{-5}c.\label{cvb}\ee
It defines the parameter space associated with the parameters $b$ and $c$, depicted below:

\begin{figure}
\begin{center}
\includegraphics[width=1.9in]{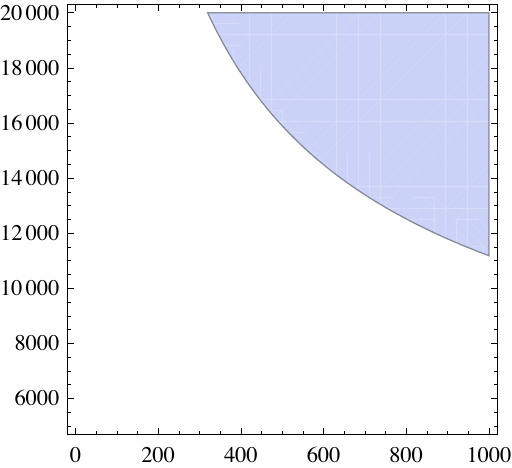}
\end{center}
\caption{\small The parameter space associated to the parameters $b$ and $c$, where $4.8 \times 10^3\leq c \leq 2\times 10^4$ and $0\leq b\leq 10^3$.}
\label{fig.1}
\end{figure}
In Figure 1 the gray region of the parameter space $c\times b$, given by (\ref{cvb}) allows a well defined background.  Note that the viability of the background, elicited by constraining the parameters via the study of the monopole solution, is far from trivial. It illustrates the importance of searching for physical systems whose typical scale is changed in a non factorable background.
{
\section{Final Remarks}

The typical radius of a monopole living in a thick brane is shown to be dependent on the position in the brane. The naked radius is dressed by the warp factor due to the monopole radius scales as the inverse of its mass. Such mass is dressed by the warp factor via the vector boson mass, arising in the spontaneous symmetry breaking. Since the monopole radius becomes dependent on the monopole position within the brane, it is possible to use it, in order to impose an upper limit on the brane thickness.

We remark that the typical constraint imposed by the upper limit in the brane thickness is compatible with further explorations in the context of thick braneworlds. For instance, studying the mass hierarchy problem in the thick brane context, there is a somewhat artificial procedure to add a  thin probe brane with positive tension,  some distance away from the thick brane \cite{AR}. Such procedure is not completely rigorous, since the gravitational probe brane contribution to the background is not taken into account. Furthermore, the trapping of matter into the probe brane is demanded. Meanwhile this procedure implies that any mass parameter measured on the probe brane scales as (see Eq. (\ref{13})) $m_{probe}=m/[2\cosh(cr_{0})]\el{b}$, where $r_{0}$ denotes the probe brane position, and the background is fixed by Eq. (\ref{20}). In order to reproduce TeV scales from Planckian scales, the condition $[2\cosh(cr_{0})]\sim 10\el{16}$ must hold. Therefore, since $r_{0}>r\el{*}$, the constraint (\ref{23}) is not necessarily violated. The new parameter $r_{0}$ may be suitably chosen, in order to accommodate both requirements.

Figure 1 regards the experimental and theoretical constraints \cite{kapp}, together with the conditions (\ref{23}) and (\ref{cvb}). It evinces the parameter space $b\times c$ in a complete, not stringent, analysis on the parameters related to the $AdS$ curvature and the brane thickness as well.

{We also investigated some monopole aspects when dealing with the gauge field localization. As the gauge field zero mode appears as a constant in the extra dimension it is not expected any modification in the Higgs potential. However, as evinced by Eq. (\ref{ULT}), the model depended procedure may render important quantum corrections.}

Finally, we want to emphasize that the main features concerning the monopole radius behavior can be shared by any classical, and stable, field solution with internal structure. In order to illustrate it,  a standard Nielsen-Olesen string \cite{NE} living in the thick brane is briefly analyzed. The typical effective action is given by
\ba S_{eff}&=&\int d\el{4}x \sqrt{-\bar{g}}\left( -\frac{1}{4}\bar{g}\el{\mu\alpha}\bar{g}\el{\nu\beta}F_{\mu\nu}F_{\alpha\beta}\right. \nonumber\\&&\left.+\frac{1}{2}\bar{g}\el{\mu\nu}D_{\mu}\phi (D_{\nu}\phi)\el{*}+c_{2}|\phi|\el{2}-c_{4}|\phi|\el{4}\right)\label{fr1}\ea  where $D_{\mu}=\partial_{\mu}+ieA_{\mu}$ and $F_{\mu\nu}=\partial_{[\mu}A_{\nu]}$. Hence, after the wave function renormalization $\phi \mapsto e\el{-A}\phi$, it reads\ba \hspace*{-0.2cm}S_{eff}=\int d\el{4}x \Bigg\{ -\frac{1}{4}F_{\mu\nu}F^{\mu\nu}+\frac{1}{2}|D_{\mu}\phi|\el{2}+\tilde{c}_{2}|\phi|\el{2}-c_{4}|\phi|\el{4}\Bigg\},\nonumber\ea where $\tilde{c}_{2}=e\el{2A}c_{2}$ is the dressed parameter. In the string configuration, there are two  characteristic lengths depending on  $\tilde{c}_{2}$. The first one given by $\tilde{\lambda}=\sqrt{2c_{4}/e\el{2}\tilde{c}_{2}}=e\el{-A}\lambda$ measures the region over which the field $|{\bf H}|=\frac{1}{r}\frac{d(r|{\bf A}|)}{dr}$ does not equal zero; the second characteristic length $\tilde{\xi}=1/\sqrt{2\tilde{c}_{2}}=e\el{-A}\xi$ measures the distance necessary for $|\phi|$ to approach its vacuum value. It is quite clear that the characteristic parameters behave according Eq. (\ref{21}). The interesting and prominent point is that, in the low energy regime, the Higgs type Lagrangian allows a string-like solution only if the string transverse length respects the relation $\tilde{\lambda}\sim \tilde{\xi}$ . It is obviously unaffected by the dressing factor. On the another hand, in the strong coupling limit --- when $\tilde{\lambda}$ becomes much bigger than $\tilde{\xi}$ (now interpreted as the inner core) --- these lengths must be small when compared to $\sqrt{\alpha^\prime}$. Here $\alpha^\prime$ denotes the universal slope for the state of the string with two ends. Since $\alpha^\prime\sim c_{4}/\tilde{c}_{2}$ \cite{NE}, the necessary relations guaranteeing thin strings as a good approximation, namely $\sqrt{\alpha^\prime}/\tilde{\lambda} \gg 1$ and $\sqrt{\alpha^\prime}/\tilde{\xi} \gg 1$, are the same as that imposed on the naked parameters.

\acknowledgments
The Authors thank to Prof. M. B. Hott for useful discussions  and to Prof. V. Folomeev, for bringing Ref. \cite{AR} to our attention. R. da Rocha is grateful to Conselho Nacional de Desenvolvimento Cient\'{\i}fico e Tecnol\'ogico (CNPq) grants CNPq 476580/2010-2 and 304862/2009-6 for financial support. JMHS is grateful to CNPq grant 482043/2011-3.\appendix \section{Why thick branes?}
In a thin braneworld, the Planck scale is suppressed by the warp factor on the visible brane, generating TeV scales, desirable for the solution of the hierarchy problem. However, violating barionic (and leptonic) number vertex operators
need a small coupling constant in order to have mensurable effects merely at high energies, of Planck order. 
The suppression of Planck scales on the visible brane, in consonance with Ref. \cite{AH}, is provided, given a typical leptonic number violation vertex operator in five dimensions $ S\sim \int d\el{5}x \sqrt{-G} (QQQL)$, taking into account  the quark $Q$ and lepton $L$  five-dimensional `wave functions', living in the thick brane. The extra-dimensional dependence gives \cite{AH} $Q\sim e\el{-\rho\el{2}r\el{2}}q(x\el{\mu})$ and $L\sim e\el{-\rho\el{2}(r-\mathring{r})}l(x\el{\mu}),$ where $\rho$ is a parameter of order of the four-dimensional fermionic mass scale (see \cite{AH} for the details) and $\mathring{r}$ is responsible to stuck the wave functions for quarks and leptons at different locations on the brane. One may suppose a Gaussian warp factor as $\exp{(-2\tau\el{2}r\el{2})}$, with $[\tau]=(length)\el{-2}$,  to mimic a thick brane behavior. Then \ba S&\sim& 
 N e^{(\rho\el{2}\mathring{r}\el{2}((-1+\frac{\rho\el{2}}{\tau\el{2}+\rho\el{2}}))} \int d\el{4}x (qqql),\nonumber\label{ap5} \ea where $N$ denotes the Gaussian integration. Note that the effective coupling constant associated to the danger vertex is always a negative exponential, solving the problem of a barionic number violation. If the brane has no thickness, then $\mathring{r}=0$ and the effective coupling constant is $1$, showing a cumbersome situation.
 \section{{The gauge field localization issue}}
{}{
It is important to present few remarks concerning our approach to deal with magnetic monopoles within the thick brane. More specifically, we shall comment on the (non-abelian) gauge field localization and its related consequences to our problem.  For our purposes, once the gauge field is localized, the discussion of the previous Sections applies approximately in the same way. It is worthwhile, however, to call attention on the details and differences about considering the localization of the gauge field.}

{}{As it is well known, the gauge field is somewhat claustrophobic, making its localization a difficult issue. Here, we shall use the smearing out functional approach developed in Ref.} \cite{SME}. {}{In this context, the gauge
field localization is accomplished by the introduction of a function $G(\bar{\phi})$, where  $G(\phi)$ is assumed to be a function of the minimum energy solution, $\bar{\phi}(r)$ --- representing the brane, such that there is no contribution of the gauge field zero-mode to the energy of the system.}

{}{In order to localize the gauge fields we start from its five-dimensional action }\ba
\hspace*{-.3cm}{S_{F_{MN}^a} }&=&\left. -\frac{1}{4}\int d^5x\sqrt{g}G(\bar{\phi})F_{MN}F^{MN}\right.\nonumber\\&
=&\left. -\frac{1}{4}\!\!\left(\int_{-\infty}^{+\infty}\alpha^2 G(\bar{\phi})dr\right)\!\!\!\int d^4x F_{\mu\nu}F^{\mu\nu}\right.\label{vqv}
\ea {}{Notice that by decomposing the five-dimensional gauge field in the usual way, i. e., $A_{\mu }=\sum_{n}a_{\mu }(x)\alpha _{n}(r)$, the coefficients $\alpha_{n}$ remain the same as the standard (abelian) case. Therefore, the equation of motion, in the gauge $\partial^{\rho }A_{\rho}=0$ and $A_{4}=0$, reads} \begin{equation}
m_{n}^{2}\alpha _{n}(r)+e^{2A}\Big[\alpha _{n}^{\prime \prime }(r)+\Big(%
\frac{G^{\prime }(\bar{\phi})}{G(\bar{\phi})}+2A^{\prime }\Big)\alpha
_{n}^{\prime }(r)\Big]=0.  \label{11}
\end{equation}%
{}{By introducing}
$
\alpha _{n}(r)=e^{-\gamma (r)}g_{n}(r),
$ {and setting $2\gamma ^{\prime }=2A^{\prime }+G^{\prime
}/G $, it results in the Schr\"odinger
equation:}%
\begin{equation}
-g_{n}^{\prime \prime }(r)+\big[\gamma ^{\prime \prime }+(\gamma ^{\prime
})^{2}-m_{n}^{2}e^{-2A}\big]g_{n}(r)=0.  \label{131}
\end{equation}%
{}{For the massless zero mode $(g_{0}\equiv g)$ one has} \cite{SME}
\ba-g^{\prime \prime }(r)+[\gamma ^{\prime \prime }+(\gamma ^{\prime
})^{2}]g(r)=0\nonumber\ea {}{and therefore $g(r)\sim e^{\gamma (r)}$. In this was $\alpha _{0}$ ends up as a
constant.} {}{Indeed, the zero mode gauge field permeates the whole bulk as a constant. In addition, it is localized by means of a normalizable (in the entire extra-dimensional domain) smearing out function, presenting a narrow bell shape profile on the brane. It is performed in such a way that the factor $\int_{-\infty}^{+\infty}\alpha^2 G(\bar{\phi})dr $ introduced in Eq.} (\ref{vqv}) {}{equals the unity. In this vein, we do not expect any effect of the gauge field localization in the Higgs potential and all our previous assertions concerning the monopole issue hold.}

{}{However, as it is clear from the  procedure adopted, the normalization is indeed dependent of the model. Hence, it would be quite interesting to explore a little further what type of modification accrue in considering the localization of the gauge fields. As we shall see, there is a subtle, although very important, consequence. By carring out a similar procedure to the non-abelian case, and denoting} \ba
\kappa^2 = \int_{-\infty}^{+\infty}\alpha^2 G(\bar{\phi})dr\label{kappa}
\ea{}{ it follows that the effective action reads now} $S_{eff}=\int \mathcal{L}\;d\el{4}x$, where \ba \hspace{-.1cm}\mathcal{L}=-\frac{\kappa^2}{4}F_{\mu\nu}\el{a}F\el{a\mu\nu}+\frac{e\el{2A}}{2}D_{\mu}\phi\el{a}D\el{\mu}\phi\el{a}-\frac{\lambda}{8} e\el{4A}\Big(\phi\el{a}\phi\el{a}-v\el{2}\Big)\el{2}.\nonumber\label{51}\ea {}{Therefore, the $\kappa^{2}$ factor may be absorbed in the gauge field rescaling} \ba W_{\mu}\el{b}\mapsto \kappa^{-1}W_{\mu}\el{b}\qquad{\rm and}\qquad W^{\mu b}\mapsto \kappa^{-1}W^{\mu b}.
\ea {}{Taking into account Eqs.}(\ref{3}) {}{and} (\ref{4}) {}{rescaled according to the prescription above, such a rescaling implies the following modifications in three terms coming from the explicit expression for $\kappa^2F_{\mu\nu}\el{a}F\el{a\mu\nu}$ in  $\mathcal{L}$  above:}
\ba
\kappa^2\left(\partial_{[\mu}W_{\nu]}\el{a}\partial^{[\mu}W^{\nu]a}\right)&\mapsto&\left(\partial_{[\mu}W_{\nu]}\el{a}\partial^{[\mu}W^{\nu]a}\right),\nonumber\\
-2\kappa^2e\varepsilon\el{a}_{bc}\partial_{[\mu}W_{\nu]}W_{\mu}\el{b}W_{\nu}\el{c}  &\mapsto&-2\kappa^{-1}e\varepsilon\el{a}_{bc}\partial_{[\mu}W_{\nu]}W_{\mu}\el{b}W_{\nu}\el{c}, \nonumber\\
e^2\kappa^2\,\varepsilon\el{a}_{bc}\varepsilon\el{a}_{rs}W_{\mu}\el{b}W_{\nu}\el{c}
W^{\mu r}W^{\nu s} &\mapsto& e^2\kappa^{-2}\,\varepsilon\el{a}_{bc}\varepsilon\el{a}_{rs}W_{\mu}\el{b}W_{\nu}\el{c}
W^{\mu r}W^{\nu s}\!\! 
\nonumber
 \ea\noindent 
{}{Therefore, the coupling constant $e$ must be replaced by $\mathring{e} = e\kappa^{-1}$. Note that the replacement of $e$ by $\mathring{e}$ is enough for the scalar field kinetic term. Hence, it follows that $M_{W}=\mathring{e}\tilde{v}$ and the monopole mass goes as $\tilde{M}_{m} = \kappa e^{A}M_{m}$. Therefore we arrive at the same qualitative behavior obtained without the parameter $\kappa$. Nevertheless, as may be readily verified, the monopole radius rescaling is such that the product $\tilde{R}_{m}\tilde{M}_{m}$ is no longer invariant, rendering the equality} \be{ \tilde{\lambda}_{c}}/{\tilde{R}_{m}}=\kappa^{-3}{\lambda_{c}}/{R_{m}},\label{ULT}\ee {}{where, as before, $\lambda_{c}$ is the monopole's Compton wavelength. In other words, the localization procedure indicates that in some backgrounds, for which $\kappa \ll 1$, $\tilde{\lambda}_{c}\sim \tilde{R}_{m}$ i. e., the quantum correction may be in order.}}


\begin{thebibliography}{0}

\bibitem{RSI} L. Randall and R. Sundrum, Phys. Rev. Lett. {\bf 83}, 3370 (1999) [{\tt arXiv:hep-ph/9905221}].

\bibitem{kapp} D. J. Kapner et al., Phys. Rev. Lett. {\bf 98}, 021101 (2007); R. Casadio, O. Micu, Phys. Rev. D {\bf 81}, 104024 (2010).

\bibitem{MG} M. Gremm, Phys. Lett. B {\bf 478}, 434 (2000).

\bibitem{OT} O. De Wolf, D. Z. Freedman, S. S. Gubser, and A. Karch, Phys. Rev. D {\bf  62}, 046008 (2000); C. Csaki, J. Erlich, T. Hollowood, and Y. Shirman, Nucl. Phys. B {\bf  581}, 309 (2000).

\bibitem{VVM} V. Dzhunushaliev, V. Folomeev, and M. Minamitsuji, [{\tt arXiv:0904.1775 [gr-qc]}].

\bibitem{TP} G. 't Hooft, Nucl. Phys. B {\bf 79}, 276 (1974); A. M. Polyakov, JETP Lett. {\bf 20}, 194 (1974).

\bibitem{DIRAC} P. A. M. Dirac, Proc. Roy. Soc. A {\bf 133}, 60 (1931); P. A. M. Dirac, Phys. Rev. {\bf 47}, 817 (1948).

\bibitem{NS} P. Goddard and D. Olive, Rep. Prog. Phys. {\bf 41}, 1357 (1978).

\bibitem{NSN} T. W. Kirkman and C. K. Zachos, Phys. Rev. D {\bf 24}, 999 (1981).

\bibitem{JP} J. Preskill, Ann. Rev. Nucl. Part. Sci. {\bf 34}, 461 (1984).

\bibitem{SUP} M. Cvetic, S. Griffies, and S. Rey, Nucl. Phys. B {\bf 381}, 301 (1992); K. Skenderis and P. K. Towsend, Phys. Lett. B {\bf 468}, 46 (1999).

\bibitem{WEI} H. Georgi, H. Quinn, and S. Weinberg, Phys. Rev. Let. {\bf 33}, 451 (1974).

\bibitem{SME} A. E. R. Chumbes, J. M. Hoff da Silva and M. B. Hott, \emph{Phys. Rev. D} {\bf 85}, 085003 (2012) [{\tt arXiv:1108.3821 [hep-th]}]].

\bibitem{AR} N. Barbosa-Cendejas, A. Herrera-Aguilar, K. Kanakoglou, U. Nucamedi, and I. Quiros, [{\tt arXiv:0712.3098 [hep-th]}].

\bibitem{NE} H. B. Nielsen and P. Olesen, Nucl. Phys. B {\bf 61}, 45 (1973).

\bibitem{AH} N. Arkani-Hamed and M. Schmaltz, Phys. Rev. D {\bf 61}, 033005 (2000).

\end{thebibliography}
\end{document}